\documentclass[conference]{IEEEtran}
\IEEEoverridecommandlockouts
\usepackage{cite}
\usepackage{amsmath,amssymb,amsfonts}
\usepackage{algorithmic}
\usepackage{graphicx}
\usepackage{textcomp}
\usepackage{xcolor}
\usepackage{placeins}
\usepackage{subcaption}

\newcommand{\R}[0]{\mathbb{R}} 
\newcommand{\C}[0]{\mathbb{C}} 

\newtheorem{lemma}{Lemma}
\newtheorem{proposition}{Proposition}

\newtheorem{definition}{Definition}

\usepackage{todonotes} 

\def\BibTeX{{\rm B\kern-.05em{\sc i\kern-.025em b}\kern-.08em
    T\kern-.1667em\lower.7ex\hbox{E}\kern-.125emX}}
\begin{document}

\title{Fast-Fourier-Forecasting Resource Utilisation in Distributed Systems\\
\thanks{
Paul Pritz acknowledges the financial support by the Computing Department at Imperial College London.
Thanks are also due to Tiffany Tuor of Imperial College London and Shiqiang Wang of IBM Research for their insightful discussions.}
}

\author{

\IEEEauthorblockN{Paul J. Pritz}
\IEEEauthorblockA{\textit{Department of Computing} \\
\textit{Imperial College London}\\
London, United Kingdom \\
paul.pritz18@imperial.ac.uk}
\and

\IEEEauthorblockN{Daniel Perez}
\IEEEauthorblockA{\textit{Department of Computing} \\
\textit{Imperial College London}\\
London, United Kingdom \\
daniel.perez@imperial.ac.uk}
\and

\IEEEauthorblockN{Kin K. Leung}
\IEEEauthorblockA{\textit{Department of Computing} \\
\textit{Imperial College London}\\
London, United Kingdom \\
kin.leung@imperial.ac.uk}
\and

}

\maketitle

\begin{abstract}
Distributed computing systems often consist of hundreds of nodes (machines), executing tasks with different resource requirements.
Efficient resource provisioning and task scheduling in such systems are non-trivial and require close monitoring and accurate forecasting of the state of the system, specifically resource utilisation at its constituent machines.
Two challenges present themselves towards these objectives. \newline
First, collecting monitoring data entails substantial communication overhead.
This overhead can be prohibitively high, especially in networks where bandwidth is limited.
Second, forecasting models to predict resource utilisation should be accurate and also need to exhibit high inference speed.
Mission critical scheduling and resource allocation algorithms use these predictions and rely on their immediate availability. \newline
To address the first challenge, we present a communication-efficient data collection mechanism.
Resource utilisation data is collected at the individual machines in the system and transmitted to a central controller in batches.
Each batch is processed by an adaptive data-reduction algorithm based on Fourier transforms and truncation in the frequency domain.
We show that the proposed mechanism leads to a significant reduction in communication overhead while incurring only minimal error and adhering to accuracy guarantees.
To address the second challenge, we propose a deep learning architecture using complex Gated Recurrent Units to forecast resource utilisation.
This architecture is directly integrated with the above data collection mechanism to improve inference speed of the presented forecasting model.
Using two real-world datasets, we demonstrate the effectiveness of our approach, both in terms of forecasting accuracy and inference speed. \newline
Our approach resolves several challenges encountered in resource provisioning frameworks and can also be generically applied to other forecasting problems.

\end{abstract}


\section{Introduction}
\label{sec:intro}
Distributed systems usually consist of hundreds or thousands of nodes (machines).
Efficient management of such systems is often challenging and requires collecting and forecasting of the utilisation of resources such as CPU and memory of the machines of the system.
In practice, resource over and under provisioning are common and overall resource utilisation is often poor, leading to a waste of computational resources \cite{analysis_of_alibaba_traces_2019}.
Data collection in some distributed systems is further hindered by communication constraints, especially in systems that are not interconnected by a high-bandwidth network.
For instance, sensor networks or networks involving edge devices may suffer from significant communication constraints.
This leads to two concrete challenges as follows.
First, a central controller that manages scheduling and resource allocation needs to collect monitoring data from all nodes in the network in a communication-efficient manner.
Second, the scheduler requires an efficient forecasting model.
Efficiency in this context encompasses both accuracy as well as inference speed.

Due to the aforementioned communication constraints it is often detrimental to send all of the collected data to the central controller, forcing the local machines to reduce or compress data before transmission.
To address this challenge, we present a data reduction mechanism based on Fourier transforms that can significantly reduce the communication overhead of transmitting monitoring data.
Our experiments on real-world data in Section \ref{sec:experiments} demonstrate that communication savings in excess of 60\% can be achieved while only incurring minimal error in the transmitted data and achieving prediction accuracy comparable to our benchmark model.
The proposed methodology can be combined with lossless compression algorithms and exhibits error bounds, which we derive in Section  \ref{method:fourier_truncation}.

As a forecasting model, we propose the use of a deep learning architecture, based on complex Gated Recurrent Units (cGRU).
In real-world systems, deep learning models are often not practical since training of such models and their use for inference tends to be computationally expensive.
The data collection mechanism we present in this paper, however, can be directly combined with our proposed forecasting architecture to improve both training and inference speed, which we demonstrate in Section \ref{sec:experiments}, using real-world datasets.

The methods presented in this paper are developed specifically with distributed systems in mind.
However, the individual components can be easily applied to different time series forecasting problems.
The integration of the proposed data processing and inference techniques with scheduling and further system management is left for future work.
Our main contributions are as follows:
\begin{enumerate}
    \item We propose a communication-efficient algorithm for time series data transmission in distributed systems, using a batched data transfer protocol with a Fourier transform based mechanism for data reduction.
    \item We show that our data transmission algorithm can be readily applied to improve the inference speed of recurrent neural networks, specifically complex Gated Recurrent Units.
    \item We propose a deep learning architecture for forecasting resource utilisation in distributed systems and conduct extensive experiments using real-world datasets that demonstrate the effectiveness of our proposed methodology.
\end{enumerate}

The remainder of this paper is structured as follows. 
In Section \ref{sec:literature}, we discuss existing literature.
The proposed methodology is presented in Section \ref{sec:method}, starting with the data transmission protocol before introducing the proposed Fourier processing mechanism and forecasting model.
The experiments using real-world datasets are discussed in Section \ref{sec:experiments}.
Lastly, we conclude in Section \ref{sec:discussion} and provide directions for future research.

\section{Related Work}
\label{sec:literature}
Previous literature relevant to the approaches outlined in Section \ref{sec:method} can be broadly categorised into the areas of resource utilisation forecasting, data collection in distributed systems and complex valued recurrent neural networks.

Ample previous research studies the use of classical time series models, such as autoregressive and moving average models for load forecasting, while some have also explored neural networks and alternative models such as support vector regression.
A comprehensive overview of previously studied forecasting models for cloud workloads is provided by \cite{Vzquez_time_series_forecasting_2015}. 
The authors of \cite{ren_dynamic_load_balancing_2011}, \cite{Chandra_dynamic_resource_alloc_2003} and \cite{efficient_autoscaling_2011} propose the use of classical time series models.
\cite{Chandra_dynamic_resource_alloc_2003} and \cite{efficient_autoscaling_2011} both propose autoregressive models for load forecasting, using a simple autoregressive model and ARMA models (autoregressive moving average) respectively.
\cite{ren_dynamic_load_balancing_2011} propose a load balancing algorithm for cloud infrastructures, as part of which they employ an exponential smoothing based forecasting method.
\cite{ISLAM_empirical_model_load_2012}, \cite{tuor_online_collection_2019} evaluate neural network based approaches.
\cite{ISLAM_empirical_model_load_2012} evaluate several neural network architectures and compare them against linear regression using simulated data.
The approach proposed by \cite{tuor_online_collection_2019} combines a threshold-based method for communication reduction in collecting utilisation data in distributed systems and couples this with a k-means clustering where a forecasting model predicts the centroid values to infer resource utilisation values of individual machines.
Alternative modelling approaches including support vector regression, Markov chain models and exponential smoothing based models are presented by \cite{efficient_resource_provisioning_hu_2013}, \cite{Zhong_pso_WWSVM_2018}, \cite{gong_press_2010} and \cite{huang_double_exp_smoothing_2012}.
Several previous papers use the same datasets we use to evaluate their models, enabling comparison among the different approaches \cite{efficient_resource_provisioning_hu_2013, Zhong_pso_WWSVM_2018, tuor_online_collection_2019, gong_press_2010}.

Further to load forecasting models, several approaches for reducing the communication overhead in distributed systems have been proposed in the body of existing literature.
A number of existing methods use a set of randomly selected nodes in the system to infer data for the remaining unobserved nodes using matrix completion \cite{compressed_acquisition_2014, correlation_estimators_dist_source_coding_2012, lossy_compression_2011, anagnostopoulos_matrix_completion_2014}.
The approaches in \cite{online_method_for_minimizing_monitoring_2015, krause_opt_sensor_placement_2008} also use a set of randomly selected nodes, but employ Gaussian methods to infer unobserved data.
For both approaches, data is only being collected for a random subset of nodes, which may lead to resource utilisation imbalances and deviations in accuracy between different nodes.
Other algorithms, relying on a per-node condition, i.e. avoiding the problems of imbalance, are presented in \cite{ARIMA_communication_sensor_liu_2005, adaptive_sampling_sensornet_law_2009, adaptive_data_collection_harb_2016, sensor_sampling_frequency_chatterja_2008, wind_sensor_sampling_2018, guitton_fourier_2007}.
The proposed methods apart from the ones in \cite{wind_sensor_sampling_2018} and \cite{guitton_fourier_2007} do not use Fourier transforms for data reduction.
Both approaches rely on heuristics and do not use the reduced data by Fourier processing as a means to accelerate inference or model training at a central controller.
While there is little research on the use of Fourier transforms for data collection in distributed systems, previous research has explored its uses for correlation approximation and similarity search in databases that has some similarities to the idea we propose \cite{DFT_phd_thesis_2004, similarity_search_1993, fast_approx_correlation_2010}.

The methods in \cite{FCNN_pratt_2017} and  \cite{fourier_rnn_wolter_2018} explore the combination of Fourier transforms and deep learning.
Specifically, \cite{FCNN_pratt_2017} propose the computation of convolutions in the frequency domain to speed up the training of convolutional neural networks and the approach in \cite{fourier_rnn_wolter_2018} uses windowing to reduce the training time of recurrent neural networks where Fourier transforms serve as a pre-processing step.

Our work adds to the body of existing research by introducing a Fourier transform based methodology that can be used to reduce communication overhead and directly accelerate the training of recurrent neural networks.

\section{Proposed Methodology}
\label{sec:method}
Our proposed methodology consists of three key components, namely a batched data collection algorithm, a Fourier transform mechanism for data reduction and a deep learning model that can leverage the former two components to achieve improvements in inference speed.

\subsection{Batched Data Collection}
\label{method:data_collection}
Let $M := \{M_1,M_2,...,M_p\}$ be a set of $p$ nodes in a distributed system, all connected to a network that allows them to communicate with a central controller $C$.
We assume that time is slotted and that the nodes $M$ observe data at pre-defined, discrete time steps.
To manage the nodes $M$ and schedule computational jobs, $C$ requires information about the nodes' resource utilisation.
The individual nodes collect their respective resource utilisation data locally.
These form a discrete-time time series $x_{M_i}$ for each node $M_i \in M$.
New observations are appended to this series as data is collected.
Two protocols for data transmission are readily conceivable.
Each node $M_i$ could decide whether to transmit an observation at the time it is collected.
The central controller would then treat the last received data point as the current state of node $M_i$~--~potentially interpolating if no update has been received for some time.
\cite{tuor_online_collection_2019} introduces such a methodology, where the nodes in a distributed system make an update decision at each time step.
Alternatively, each node $M_i$ can collect its respective time series locally for a given number of time steps and then send it to $C$ in one batch.
The central controller then has to estimate the current states of all nodes $M$ in time steps between consecutive updates using some forecasting model or method of interpolation.
This batched processing approach~--~which our proposed methodology relies on~--~will be referred to as a \textit{batched data collection mechanism}.

Before generalising to the complete set of nodes $M$, consider a single node $M_i$.
Let $T$ be a set of discrete time steps at which the variable of interest is observed and define the time between two time steps as $\tau = t_j - t_i, j = i+1$. 
Let $\theta$ denote the update times where a batch update is sent to the central controller.
The time between two consecutive updates is then defined as $n \times \tau := \theta_j - \theta_i, j = i + 1, 2 \leq n$.
The number of time steps $n$ between batch update times is kept constant, i.e., the batch update times are equally spaced.
This is equivalent to saying: The central node $C$ will receive a batch update every $n$ time steps.
\begin{figure}
    \centering
    \includegraphics[width=1\linewidth]{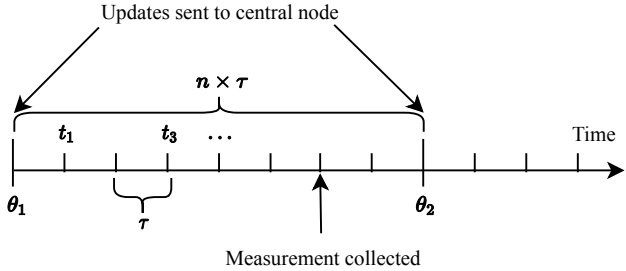}
    \setlength{\belowcaptionskip}{-10pt}
    \caption{Data collection and update procedure for a single node $M_i$. Measurements are collected at the discrete time steps $t$ and and the central node $C$ is updated every $n$ time steps, i.e., at update times $\theta$.}
    \label{fig:data_update_timeline}
\end{figure}
This batched update mechanism is illustrated in Fig. \ref{fig:data_update_timeline}.
At each update step, the node $M_i$ will have collected $n$ observations, forming a time series $u \in \R^n$.
Each observation measurement in $u$ is represented as a floating-point number.
To achieve any communication savings, fewer than $n$ floating-point numbers must be transmitted to the central controller by reducing the data at the individual nodes.
We propose an algorithm for achieving such data reduction in the next sub-section.

\subsection{Fourier Truncation}
\label{method:fourier_truncation}
The idea behind the proposed data-reduction mechanism is as follows.
A time series can be converted to its frequency-domain representation using Fourier transforms.
If only a few terms of the Fourier transforms are sufficient to capture the majority of the variation of the time series then it suffices to transmit these few terms to the central controller.
This way, the data batch as defined in Section \ref{method:data_collection} can be reduced to fewer than $n$ floating point numbers.
We start by giving the necessary definitions for the proposed methodology, before describing two approaches for choosing the number of terms to include.
The Fourier transform of the discrete time-domain signal or time series $u$ is denoted by $U$ and given by
\begin{equation}
    U_f = \sum_{i=0}^{n-1} u_i e^{ -j (2\pi / n)if} \,.
    \label{eq:DFT_of_u}
\end{equation}
$U \in \C^n$ is a sequence of complex numbers of the same length as $u$.
The Fourier transform of a real-valued time series, $u := u_0, u_1, ... , u_{n-1}$ of length $n$, where $u_i \in \R, 0 \leq i < n$, has the useful property of complex conjugacy, s.t.
\begin{equation}
    u_i = u_{-i}^* \,,
    \label{eq:compl_conj}
\end{equation}
where $*$ denotes the complex conjugate.
Exploiting this property, only $n/2+1$ terms of the Fourier transforms of a real-valued time series are required to fully capture the series.
Since the resource utilisation data under consideration is real-valued, we can directly use this property to reduce the data volume of the Fourier transforms to almost the same volume as the original time-domain data~--~recall that a complex number is represented using two floating point numbers. 

To further reduce the amount of data in the frequency domain, we define a methodology referred to as Fourier truncation, which is equivalent to an adaptive low-pass filter without attenuation.
\begin{definition}
    The energy of the discrete-time signal $u$ with length $n$ is given by
    \begin{equation}
        E(u) = \sum_{i=0}^{n-1} |u_i|^2 \,.
    \end{equation}
    \label{def:signal_energy}
\end{definition}
\begin{lemma}
    Using Parseval's theorem and Definition \ref{def:signal_energy}, the energy of a signal is preserved after the Fourier transform according to
    \begin{equation}
        E(u) = \sum_{i=0}^{n-1} |u_i|^2 = \frac{1}{n} \sum_{f=0}^{n-1} |U_f|^2 \,,
        \label{eq:parseval_theroem}
    \end{equation}
    where $U_f$ is the Fourier transform of $u_i$.
    \label{lemma:energy_equivalence}        
\end{lemma}
Lemma \ref{lemma:energy_equivalence} forms the fundamental background for our Fourier truncation methodology, since it details the relationship between signal energy in the time and frequency domain.
We propose two ways of choosing the number of terms $k$ to transmit to the central controller, one using an absolute error criterion and one using a relative similarity criterion.
\begin{definition}
    The series $R$ is the truncated version of the Fourier transform $U$ of $u$ that includes all terms up to term $k$, s.t. $R := U_{0 \leq i < k} \,, k \leq n$. 
    We also define the energy of the truncated series $R$ as $E(R)$.
    \label{def:trunc_series}
\end{definition}

The truncated series $R$ only has $k$ terms.
By truncating and exploiting the complex conjugacy property (Equation \eqref{eq:compl_conj}), the number of floating point numbers that have to be transmitted is therefore reduced by $2(n-k)$.

\begin{definition}
    As a measure of deviation between the original time series and the inverse of the truncated frequency-domain representation, we define the root mean squared error (RMSE) between the original time series $u$ and the truncated version $R$ as
    \begin{equation}
        RMSE(u, R) = \sqrt{\frac{1}{n} \sum_{i=0}^{n-1} |u_i - \mathcal{F}^{-1}(R)_i|^2} \,,
        \label{eq:RMSE_def_freq}
    \end{equation}
    where $\mathcal{F}^{-1}$ denotes the inverse discrete Fourier transform.
    We refer to this measure of error as the truncation error.
\end{definition}

Using Lemma \ref{lemma:energy_equivalence} we can impose a scale-dependent error bound in terms of the RMSE defined in Equation \eqref{eq:RMSE_def_freq}, which we detail in the following Proposition.
\begin{proposition}
    The RMSE caused by the truncation of the Fourier transform $U$ of the time series $u$ is bounded by $\epsilon_{RMSE}$ according to
    \begin{equation}
        RMSE(u, R) \leq \epsilon_{RMSE} \,,
        \label{eq:MSE_absolute_bound}
    \end{equation}
    if the number of terms $k$ to include in $R$ is chosen such that
    \begin{equation}
        \sqrt{\frac{1}{n} \Big( E(U) - E(R) \Big)} \leq \epsilon_{RMSE} \,,
        \label{eq:choose_k_for_MSE_bound}
    \end{equation}        
    where $E(R)$ is the energy of $R$ and $R$ is the truncated Fourier transform of $u$, defined as $R = U_{0 \leq i < k}$, according to Definition \ref{def:trunc_series}.
    \label{proposition:MSE_bound}
\end{proposition}
\begin{IEEEproof}
    From Definition \ref{def:trunc_series}, we define the terms of the frequency-domain representation $U$ of the time series $u$, lost in the truncation of $U$ as $L := U_{k \leq i < n}$.
    Using Definition \ref{def:signal_energy} and Lemma \ref{lemma:energy_equivalence}, the energy lost in the truncation of $U$ is given by
    \begin{equation}
        E(L) = E(U) - E(R) \,.
    \label{eq:truncated_energy_exact}
    \end{equation}
    For ease of exposition, we denote $\mathcal{F}^{-1}(R)_i$ by $r_i, 0 \leq i < n$.
    To prove Proposition \ref{proposition:MSE_bound}, we need to show that
    \begin{equation}
        RMSE(u, R) \leq \sqrt{\frac{1}{n} E(L)} \,.
        \label{eq:proof_target}
    \end{equation}
    By converting both $U$ and $R$ to the time domain and using Equation \eqref{eq:truncated_energy_exact}, this can be simplified to
    \begin{equation}
        \sum_{i=0}^{n-1} |u_i - r_i|^2 \leq \sum_{i=0}^{n-1} u_i^2 - r_i^2 \,.
        \label{eq:simple_bound}
    \end{equation}
    To prove Proposition \ref{proposition:MSE_bound}, it is therefore sufficient to show that Equation \eqref{eq:simple_bound} holds.
    Expanding the left-hand side, we get
    \begin{equation}
        \sum_{i=0}^{n-1} |u_i - r_i|^2 = \sum_{i=0}^{n-1} u_i^2 - 2 u_i r_i + r_i^2 \,.
    \end{equation}
    By subtracting the right-hand side of Equation \eqref{eq:simple_bound} and simplifying, this transforms to
    \begin{equation}
        \sum_{i=0}^{n-1} |u_i - r_i|^2 - \sum_{i=0}^{n-1} u_i^2 - r_i^2 = 2 \sum_{i=0}^{n-1} r_i (r_i - u_i) \,.
    \end{equation}
    From Lemma \ref{lemma:energy_equivalence}, we know that
    \begin{equation}
        \sum_{i=0}^{n-1} r_i^2 \leq \sum_{i=0}^{n-1} u_i^2 \iff 2 \sum_{i=0}^{n-1} (r_i + u_i)(r_i - u_i) \leq 0 \,.
    \end{equation}
    Since $u_i, r_i \geq 0$
    \begin{equation}
        2 \sum_{i=0}^{n-1} r_i (r_i - u_i) \leq 2 \sum_{i=0}^{n-1} (r_i + u_i)(r_i - u_i) \leq 0 \,,
    \end{equation}
    which proves Equation \eqref{eq:simple_bound} and therefore Proposition \ref{proposition:MSE_bound}.
    Since $E(L)$ is monotonically decreasing in $k$, $k$ can be chosen large enough such that Equation \eqref{eq:choose_k_for_MSE_bound} is satisfied.
\end{IEEEproof}

The error bound given in Proposition \ref{proposition:MSE_bound} uses a scale-dependent error metric and is thus only useful when the magnitude of the time series $u$ is known beforehand. 
To be able to generalise the methodology to arbitrary time series, a scale-independent error measure is more desirable.
We propose the use of a percentage energy threshold to be captured in the truncated time series.
We define this threshold value as $e \in [0,1]$ and use the cumulative fraction of the time series' energy captured up to each term to choose the number of terms to include given the threshold value $e$.
Let the series $S$ of cumulative sums of the energy captured in the terms of the Fourier transform $U$ of the time series $u$ be defined as
\begin{align*}
    S_0 &= 0 \\
    S_{i+1} &= S_i + E(U[i]).
\end{align*}
Further, denote the total energy of the time series or equivalently the maximum of series $S$ by $S_{max} := max(S)$.
Then the normalised series $S^{n}$ with $S_i^n \in [0,1]$, is
\begin{equation}
    S_i^n = \frac{S_i}{S_{max}} \,.
\end{equation}
The terms of $S^n$ are equivalent to the fraction of the signal's energy captured by the terms of the Fourier transform up to the $i^{th}$ term.
If instead of imposing an absolute RMSE bound the procedure based on an energy threshold value is employed, the value of $k$ is chosen such that
\begin{equation}
    E(R) = \frac{1}{n} \sum_{i=0}^{k-1} |U_i|^2 \geq E(u) \times e\,,
    \label{eq:truncated_energy_e_bound}
\end{equation} 
where $e$ is the energy threshold as defined previously.
Inequality \eqref{eq:truncated_energy_e_bound} is satisfied by choosing $k$ such that $e \leq S_k^n \land e > S_{k-1}^n$.
The Inequality \eqref{eq:truncated_energy_e_bound} can be transformed to
\begin{equation}
    \frac{E(R)}{E(u)} \geq e \,.
    \label{eq:relative_error_bound}
\end{equation}
The term on the left hand side of Equation \eqref{eq:relative_error_bound} is the similarity between the original series and the truncated version in terms of captured energy.
It can also be interpreted as the relative accuracy of the truncated version compared to the full series. 
This measure lies between 0 and 1 and is independent of the scale of the signal.
In practice it is sufficient to specify some level of minimum similarity and choose the threshold value $e$ according to Inequality \eqref{eq:relative_error_bound}.

The algorithm, resulting from the truncation methodology in combination with the proposed batched data collection is given in Fig. \ref{algo:FFT_compression_at_M}.

\begin{figure}
    \begin{algorithmic}[1]
        \STATE Initialise empty list $u$ \\
        \WHILE{$t \notin \theta$}
            \STATE Observe the variable of interest.
            \STATE Append the new observation to $u$. 
        \ENDWHILE
        \STATE $U$ = $\mathcal{F}(u)$ \\
        \STATE Choose $k$ according to either of the two proposed truncation methodologies \\
        \STATE $R = U_{0 \leq i < k}$\\
        \STATE \textbf{Transmit} $R$ to the central controller $C$ \\
    \end{algorithmic}
        \caption{Compression and data collection algorithm executed at nodes $M$}
    \label{algo:FFT_compression_at_M}
\end{figure}

Using the Fast Fourier Transform algorithm introduced in \cite{cooley_tukey_FFT_1965}, the Fourier transforms of a time series of length $n$ can be computed with a time complexity of $\mathcal{O}(n\, log(n))$.
The proposed truncation mechanisms require a complete pass of the Fourier transforms, adding another $n/2 + 1$ computation steps.
Hence, the overall number of computation steps required for a single node is $\mathcal{O}(n\, \log(n)) + n/2 + 1$, resulting in a time complexity of $\mathcal{O}(n\, \log(n))$.

\subsection{Forecasting by Complex Gated Recurrent Neural Networks}
\label{method:forecasting}

Our proposed methodology requires a forecasting model both for interpolation at time steps between consecutive batch updates as well as predicting future resource utilisation for system management purposes.
We propose the use of complex Gated Recurrent Units (cGRU) \cite{complex_GRU_wolter_2018} to forecast resource utilisation using the truncated frequency-domain representation of the resource utilisation time series that results from the data transmission methodology outlined in Section \ref{method:fourier_truncation}.
As a benchmark, we compare our approach against a time-domain GRU \cite{GRU_paper_2015}, which uses the complete time-domain representation of the input time series.
The use of gates in recurrent neural networks has been shown to improve their ability to learn longer term dependencies \cite{Hochreiter_LSTM_1997} and GRUs implement a computationally efficient gating mechanism \cite{GRU_paper_2015}.
They are therefore well suited to the problem at hand.

Throughout our experiments, we employ a sliding window model that uses a window of historic data to forecast a pre-defined number of time steps into the future.
The size of the window is defined as a multiple $w$ of the number of time steps between two successive batch updates and $l := n \times w$ is defined as the total number of observations in a window (in the time-domain).
In the time-domain, the batches in a window can simply be concatenated to form the input time series for our forecasting model. 
The problem of forecasting a pre-defined number of time steps $s$ into the future can then be defined as
\begin{equation}
    \hat{u}_{t+1}, ... , \hat{u}_{t+s} = \underset{u_{t+1}, ... , u_{t+s}}{\textrm{arg max}} p(u_{t+1}, ... , u_{t+s} | u_{t-l}, ... , u_t) \,,
    \label{eq:problem_formulation_time}
\end{equation}
where $\hat{u}_{t+1}, ... , \hat{u}_{t+s}$ are the predictions for the next $s$ time steps and $u_{t-l}, ... , u_t$ are the observations included in the input window.
The concatenation of batches to form the input window is not easily accomplished in the frequency domain due to the variable length of the truncated Fourier transforms in the batches constituting a window.
Hence, the problem formulation changes slightly to
\begin{equation}
    \hat{u}_{t+1}, ... , \hat{u}_{t+s} = \underset{u_{t+1}, ... , u_{t+s}}{\textrm{arg max}}
    p(u_{t+1}, ... , u_{t+s} | \{R_1\}, ... , \{R_w\}) \,,
    \label{eq:problem_formulation_freq}
\end{equation}
where $\{R_i\}$ is the set of all terms in the truncated frequency-domain representation of batch $i$.

Our architecture is inspired by the methodology for cGRUs using frequency-domain input proposed in \cite{fourier_rnn_wolter_2018}.
The model accepts variable length input sequences and is applied to the truncated frequency-domain representation of each batch in the input window.
The size of the model is kept constant between the time and frequency-domain models as parameters are shared for each batch in the frequency-domain window.
The hidden states of the GRU for each of the input batches are concatenated and passed to a linear layer to arrive at the frequency-domain forecasts.
An inverse Fourier transform is then applied to these to arrive at the final time-domain predictions, which are used to calculate the prediction error for backpropagation. 
Our proposed architecture can therefore be written as
\begin{align}
    x_{j_t} =& R_{j_t} \,, 1 \leq j \leq w , 0 \leq t < |R_j| \,,
    \\
    z_{j_t} =& \sigma(W_z x_{j_t} + V_z h_{t-1} + b_z) \,,
    \\
    r_{j_t} =& \sigma(W_r x_{j_t} + V_r h_{j_{t-1}} + b_r) \,,
    \\
    \begin{split}
        h_{j_t} =& \sigma(z_{j_t} \circ h_{j_{t-1}} + (1-z_{j_t}) \circ 
        \\ 
        & \phi_h (W_h x_{j_t} + V_h(r_{j_t} \circ h_{j_{t-1}}) + b_h) \,,
    \end{split}
    \\
    \hat{u}_{t+1}, ... , \hat{u}_{t+s} =& \mathcal{F}^{-1}(f(h_{1_t}, ... , h_{w_t})) \,,
\end{align}
where $\circ$ denotes the Hadamard product, $W$, $V$ and $b$ are parameter matrices and biases, $R_{j_t}$ denotes the $t^{th}$ term of the truncated Fourier transform of the $j^{th}$ batch in the window, $f(\cdot)$ is the linear layer for final prediction, $\phi(\cdot)$ denotes the hyperbolic tangent, and $z_{j_t}$ and $r_{j_t}$ are the update and reset gates respectively.
The inverse Fourier transform is denotes by $\mathcal{F}^{-1}$ and is used to convert the model outputs to the time domain.

Since the inputs $x_{j_t} \in \C$ are in the complex domain, all subsequent weight matrices and bias vectors are also complex.
However, in the spirit of \cite{fourier_rnn_wolter_2018}, it is sufficient to simply concatenate the real and imaginary parts into a vector of reals and recombine them for the inverse Fourier transform at the end of the network.
Note that contrary to the time-domain model, where the batches in the input window are concatenated along the temporal axis, the frequency-domain representations are passed as separate inputs.
Explicitly combining the frequency-domain representations would entail giving up the benefits of truncation and reduce the gains in computational overhead.
Hence, we choose to let the forecasting model implicitly learn the relationship between the different batches.

Model training is implemented using mini-batch gradient descent with a RMSprop optimiser \cite{rmsprop_hinton} and bucketisation.
Bucketisation refers to grouping series of similar lengths into one batch and zero-padding the shorter ones.
This is required to be able to use mini-bath gradient descent, which has several advantages over stochastic gradient descent, both in terms of convergence and training speed.
The prediction error is calculated using the model's time-domain predictions and the original measured data for the predicted period.
Specifically, this means that we only apply our proposed data collection mechanism to the batches in the input window, but not to the predicted time steps.
\begin{definition}
    We define the time averaged RMSE between the predicted values and the original data during the forecasting period for all machines, as
    \begin{equation}
        \overline{RMSE}(t, s) = \sqrt{\frac{1}{s \times p} \sum_{j=1}^p \sum_{i=1}^s |\hat{u}_{j_{t+i}} - u_{j_{t+i}}|^2} \,,
    \end{equation}
    where $s$ denotes the number of predicted time steps and $p$ is the number of machines.
    We refer to this error as the prediction error.
\end{definition}

While this allows the model to learn to forecast the true time series rather than the processed one, it also means that the entire unprocessed time series has to be collected for the interval used for training.
Since our methodology does not encompass model retraining, all reported communication savings etc. refer to the period after model training, where data is collected according to the methodology proposed in Sections \ref{method:data_collection} and \ref{method:fourier_truncation}.

\section{Experiment Results}
\label{sec:experiments}
\subsection{Preliminary Data Exploration}
\label{experiments:exploration}
\begin{figure}[tb]
  \begin{subfigure}{\columnwidth}
    \centering
    \includegraphics[width=0.9\linewidth]{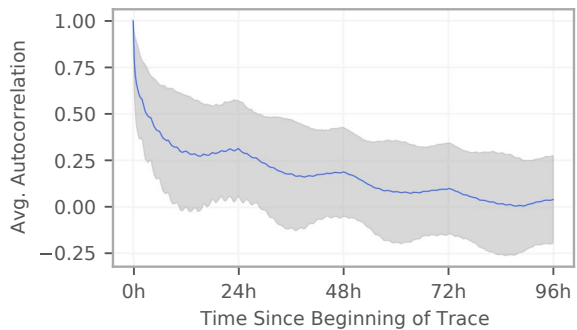}
    \caption{Google cluster}
  \end{subfigure}
  ~
  \begin{subfigure}{\columnwidth}
    \centering
    \includegraphics[width=0.9\linewidth]{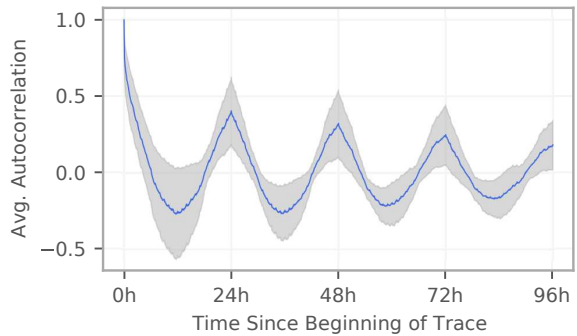}
    \caption{Alibaba cluster}
  \end{subfigure}
  \setlength{\belowcaptionskip}{-10pt}
  \caption{Average autocorrelation across all machines in the cluster traces with 95\% confidence intervals.}
    \label{fig:avg_autocorr}
\end{figure}

We evaluate the proposed methodology and forecasting models using two traces from large multi-purpose computing clusters, operated by Google and Alibaba respectively \cite{google_clusterdata_schema_2011, alibaba_trace}.
Although a resource utilisation trace generated by a sensor network would be preferable, such datasets are not currently publicly available.
Hence, we make the reasonable assumption that, given the daily cyclicality of the datasets used here, they are a good proxy for a broad range of different systems \cite{sensor_lee_2015}.
\begin{table}[b]
    \centering
    \caption{Data Summary}
    \renewcommand{\arraystretch}{1.3}
    \begin{tabular}{c|c|c}
         \textbf{Statistic} & \textbf{Google} & \textbf{Alibaba} \\
         \hline \hline
         Sampling Period & 29 days & 8 days \\
         Number of Machines & 12,480 & 4,022 \\
         Observations per Machine & 8,351 & 11,519 \\
         Sampling Frequency & 5 min & 1 min \\
         Size & 41GB & 48GB \\
         Std. Dev. (CPU) & 0.125 & 0.156 \\
         Average Utilisation (CPU) & 22.3\% & 37.4\% \\
    \end{tabular}
    \label{tab:data_summary}
\end{table}
The datasets are publicly available and have been frequently used in previous research (see for instance \cite{DeepJS_alibaba_2019, analysis_of_alibaba_traces_2019, tuor_online_collection_2019, clusterdata_Ismaeel_example2015}).
Both datasets are pre-processed to contain memory and CPU utilisation on a per machine basis for the entire sampling period.
The Google trace contains measurements for 12,480 machines over a period of 29 days, while the Alibaba trace has a sampling period of eight days and contains measurements for 4,022 machines.
\begin{figure}[tb]
  \begin{subfigure}{\columnwidth}
    \centering
    \includegraphics[width=.8\textwidth]{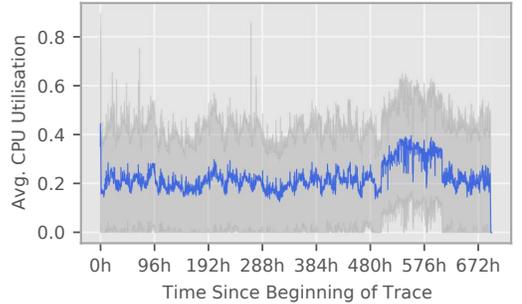}
    \caption{Google cluster}
  \end{subfigure}
  ~
  \begin{subfigure}{\columnwidth}
    \centering
    \includegraphics[width=.8\textwidth]{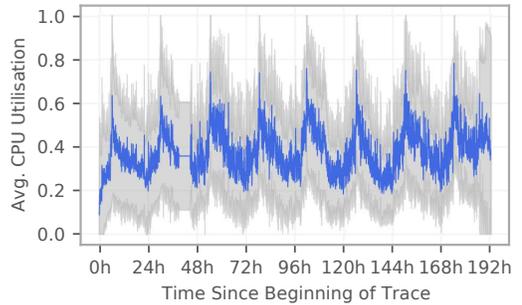}
    \caption{Alibaba cluster}
  \end{subfigure}
  \setlength{\belowcaptionskip}{-10pt}
  \caption{Average CPU utilisation across all machines in the cluster traces with 95\% confidence intervals.}
    \label{fig:cpu_avg_conf}
\end{figure}
Both clusters run a mixture of long-running services as well as batch workloads, co-hosted on the same set of machines.
Raw samples are collected in 5 minute and 1 minute intervals in the Google and Alibaba traces, respectively.
We have pre-processed the raw data to contain CPU and memory utilisation values in percent on a per machine basis, according to the methodology used in \cite{tuor_online_collection_2019}.
We also resample both traces to a sampling frequency of 5 minutes, i.e., a measurement is collected every 5 minutes, which results in a total of 8,351 and 2,302 observations per machine in the Google and Alibaba traces, respectively.
For ease of exposition, we focus our evaluation on CPU utilisation, but the same approach can be easily applied to forecast memory utilisation.
Fig. \ref{fig:cpu_avg_conf} shows the average CPU utilisation over the entire sampling period in the Google and Alibaba traces.
While both traces exhibit daily seasonality, this seasonal component is much more pronounced in the Alibaba trace than in the Google one.
This observation is also confirmed by Fig. \ref{fig:avg_autocorr}, which displays the average autocorrelation at different lags for both traces.
While both datasets exhibit fairly high autocorrelation, the Alibaba trace has a stronger seasonal (i.e., daily) component.

\subsection{Setup}

Given the daily seasonality in the data, we choose a period of 24 hours, i.e., 288 time steps as the input to our forecasting models.
As the batch length for the conducted experiments, we use a period of 6 hours, i.e., 72 data points, which results in a total of four batches per input window.
The models are trained to predict one complete batch of utilisation data, i.e., 6 hours into the future. 
This represents the minimum prediction length required to fully interpolate between the batch arrival times $\theta$.
Due to the large number of evaluation runs, we use a random sub-sample of 20 machines from each of the two datasets for both training and testing.
All models are trained on a personal computer with 32GB of RAM, an $8^{th}$ generation Intel Core i7-8700 with 3.20GHz and 6 cores and a 256GB SATA hard drive.
The truncation mechanisms from Section \ref{method:fourier_truncation} as well as the models from Section \ref{method:forecasting} are implemented in Python.
We use the PyTorch \cite{pytorch_docs} library to implement the proposed machine learning models.
The hyper parameters for each model~--~one model for each of the evaluated energy thresholds and datasets~--~are tuned using Bayesian Optimisation with an Expected Improvement acquisition function \cite{bayes_opt_OG_1998}.
Since different energy thresholds results in different data characteristics, we choose to tune the hyper parameters of each model individually to ensure optimal performance.
We split the dataset into three parts, using the first 50\% of the time steps for training, the next 25\% for hyper parameter tuning (validation) and the last 25\% for testing (i.e., prediction comparison).
We only report the results on the test set after hyper parameter tuning and complete retraining on the training and validation set.
The reported error is calculated using the model's predictions and the original dataset, i.e., without applying our proposed truncation methodology to the predicted period, but only to the model's input window.
This way of calculating the prediction error makes it more reliable as we test how well the model predicts true resource utilisation.

\subsection{Results}
We evaluate the performance of our proposed methodology in terms of communication savings, prediction errors and inference times for different energy thresholds $e$ on both the Google and Alibaba traces.

\begin{figure}[tb]
  \begin{subfigure}{\columnwidth}
    \centering
    \includegraphics[width=0.8\textwidth]{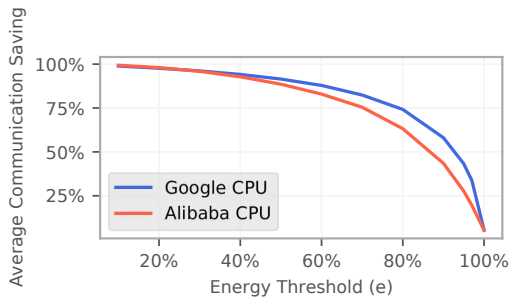}
    \caption{Communication savings at different energy thresholds}
  \end{subfigure}
  \begin{subfigure}{\columnwidth}
    \centering
    \includegraphics[width=0.8\textwidth]{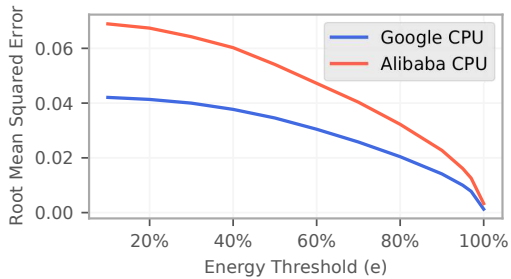}
    \caption{Error at different energy thresholds}
  \end{subfigure}
  \caption{Communication savings and truncation error (RMSE) obtained from the Fourier processing mechanism at different energy threshold values $e$ for the Google and Alibaba cluster trace sub-samples.}
  \label{fig:communication_vs_RMSE}
\end{figure}

\begin{figure}[!tb]
    \centering
    \includegraphics[width=0.8\linewidth]{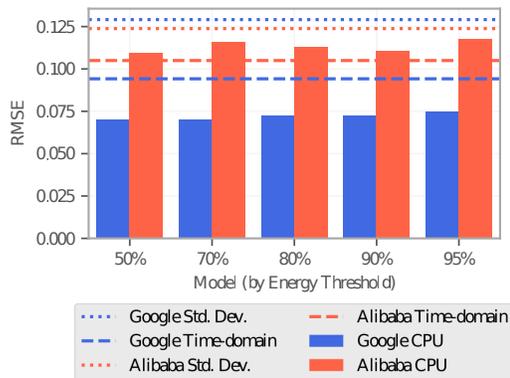}
    \setlength{\belowcaptionskip}{-10pt}
    \caption{Prediction error on the test set for models trained at different values of the energy threshold $e$ on the Google and Alibaba cluster traces.
    The time-domain benchmark models are included for both datasets.}
    \label{fig:model_losses}
\end{figure}
\begin{figure}[!tb]
    \centering
    \includegraphics[width=0.8\linewidth]{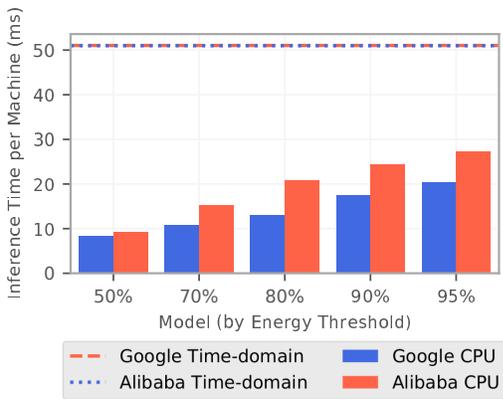}
    \setlength{\belowcaptionskip}{-10pt}
    \caption{Inference speed for a single batch for a single machine using the proposed model architecture, evaluated on the Google and Alibaba datasets at different energy threshold values $e$.
    The time-domain benchmark models are included for both datasets.}
    \label{fig:inf_speed}
\end{figure}
The Fourier processing mechanism leads to higher communication savings, the higher the error tolerance as demonstrated in Fig. \ref{fig:communication_vs_RMSE}, where error tolerance is expressed via the energy threshold.
This relationship is better than linear and communication savings in excess of 60\% can already be achieved at a small error tolerance.
From Fig. \ref{fig:communication_vs_RMSE}, it is apparent that there is a clear trade-off between the communication savings achieved via our proposed methodology and the error it introduces in the data, i.e., the truncation error.
The truncation error introduced in the data refers to the error between the original data and the truncated representation according to Equation \eqref{eq:RMSE_def_freq}, but not the prediction error.

While there is a trade-off between communication savings and error in the data, we do not find such a relationship for the prediction error.
Fig. \ref{fig:model_losses} shows the effect of different truncation thresholds on the prediction accuracy of the forecasting models.
While the Fourier processing mechanism introduces some error in the data (see Fig. \ref{fig:communication_vs_RMSE}) it also has a smoothing effect that may help to filter out some random fluctuations in the data, which could otherwise have a detrimental effect on the forecasting performance.
For the Alibaba trace, forecasting model performance in terms of the prediction error on the test set does not deteriorate significantly as the energy threshold level is reduced.
The models trained on the Google cluster trace even exhibit an improvement in performance as the energy threshold is decreased.
This improvement can be attributed to the smoothing effect of our proposed Fourier processing methodology, which reduces the noise present in the Google dataset and allows the forecasting model to learn more effectively.
Generally, the models trained on the Alibaba trace perform worse than those trained on the Google trace.
This may be due to a variety of factors, such as different hyper parameters and the shorter sampling period, leading to fewer data points in the sub-sample, and higher variance in the dataset.
The experiments on real-world data confirm that a significant improvement in inference speed can be achieved using our Fourier truncation methodology, as Fig. \ref{fig:inf_speed} demonstrates.
At an energy threshold level of $e=0.9$, the inference time of forecasting one batch of 72 time steps for a single machine is reduced by more than 50\% compared to the time-domain benchmark model.
This further decreases to less than one fifth of the time-domain model's inference time at $e=0.5$.
This improvement in inference speed is a direct result of the reduced length of the data in the input window, which entails a reduction in the amount of computations required to forecast a single batch of data.
The improvement in inference speed has two beneficial implications.
On the one hand, quick inference is often required for mission critical systems.
On the other hand, the reduction in computational overhead for forecasting resource utilisation could make deploying pre-trained models on less powerful machines in a distributed system a viable option.

\section{Conclusion and Future Work}
\label{sec:discussion}
We have proposed an approach for the efficient transmission and forecasting of time series data in distributed systems.
The approach combines a flexible data-reduction mechanism, integrated with a forecasting architecture that can achieve substantial improvements in communication overhead and inference speed.
We demonstrate the effectiveness of our approach using real-world datasets and provide a comprehensive evaluation of the proposed methodology.
Our experiments show that communication savings of approximately 60\% can already be achieved at a small error tolerance and that inference speed can be improved by more than 50\% without compromising the forecasting accuracy of our proposed model.
There are, however, some limitations to the approach that could be the subject of future research.

We have imposed error bounds to guide the data reduction rather than imposing explicit communication constraints.
While it is possible to impose explicit communication constraints, for example, by introducing an upper bound on the number of terms that can be transmitted, this would entail loosing guarantees on the error introduced by the truncation algorithm.
Extending our approach to explicitly include communication constraints is left for future work.

Our proposed truncation methodology can be described as an adaptive low-pass filter without attenuation.
Specifically, this means that we use the low-frequency terms of the Fourier transforms and eliminate some of the higher frequency terms according to the captured energy compared to the energy threshold used.
While this approach works well for the data at hand, different frequency bands may be desirable for other problems or datasets.
For instance, if the mid-frequency range captures the signal of interest, our methodology could be changed to resemble a bandpass filter.
An investigation into such an adaptation may also be a fruitful avenue for further research.
Future research can also study the integration of our data collection and forecasting methodology with scheduling and system management frameworks.

\bibliographystyle{bibliography/IEEEtran}
\bibliography{bibliography/references.bib}

\end{document}